\documentclass[journal]{IEEEtran}

\usepackage{booktabs}
\usepackage{multirow}
\usepackage{subcaption}
\usepackage{graphicx}
\usepackage[table,xcdraw]{xcolor}
\usepackage[normalem]{ulem}
\useunder{\uline}{\ul}{}
\usepackage{amsmath}

\begin{document}

\title{LSTM and GPT-2 Synthetic Speech Transfer Learning for Speaker Recognition to Overcome Data Scarcity}
%
%
%

\author{Jordan J. Bird,
Diego R. Faria,
Anik\'o Ek\'art,
Cristiano Premebida,
and Pedro P. S. Ayrosa
\thanks{J.J. Bird and D.R. Faria are with ARVIS Lab, Aston University, Birmingham, United Kingdom}
\thanks{A. Ek\'art is with the School of Engineering and Applied Science, Aston University, United Kingdom}
\thanks{C. Premebida is with the Institute of Systems and Robotics, Department of Electrical and Computer Engineering, University of Coimbra, Coimbra, Portugal}
\thanks{P.P.S. Ayrosa is with the Department of Computer Science \& LABTED, State University of Londrina, Londrina, Brazil}

}

\markboth{ArXiv Preprint - LSTM and GPT-2 Synthetic Speech Transfer Learning - JJ Bird et al.}%
{}

\maketitle

\begin{abstract}
In speech recognition problems, data scarcity often poses an issue due to the willingness of humans to provide large amounts of data for learning and classification. In this work, we take a set of 5 spoken Harvard sentences from 7 subjects and consider their MFCC attributes. Using character level LSTMs (supervised learning) and OpenAI's attention-based GPT-2 models, synthetic MFCCs are generated by learning from the data provided on a per-subject basis. A neural network is trained to classify the data against a large dataset of Flickr8k speakers and is then compared to a transfer learning network performing the same task but with an initial weight distribution dictated by learning from the synthetic data generated by the two models. The best result for all of the 7 subjects were networks that had been exposed to synthetic data, the model pre-trained with LSTM-produced data achieved the best result 3 times and the GPT-2 equivalent 5 times (since one subject had their best result from both models at a draw). Through these results, we argue that speaker classification can be improved by utilising a small amount of user data but with exposure to synthetically-generated MFCCs which then allow the networks to achieve near maximum classification scores. 
\end{abstract}


\IEEEpeerreviewmaketitle

\section{Introduction}
\begin{figure}
    \centering
    \includegraphics[scale=1.1]{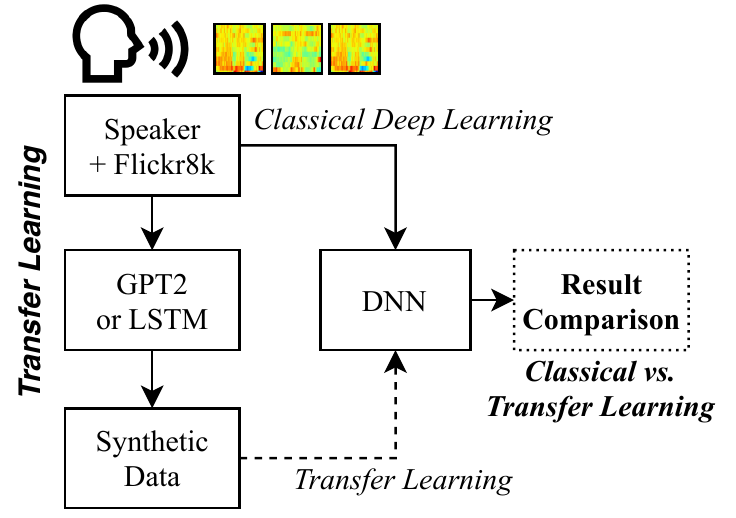}
    \caption{A simplified diagram of the experiments followed by this work towards the comparison of classical vs transfer learning from synthetic data for speaker recognition. A more detailed diagram can be found in Figure \ref{fig:method}}
    \label{fig:smalldiagram}
\end{figure}
Data scarcity is an issue that arises often outside of the lab, due to the large amount of data required for classification activities. This includes speaker classification in order to enable personalised Human-Machine (HMI) and Human-Robot Interaction (HRI), a technology growing in consumer usefulness within smart device biometric security on devices such as smartphones and tablets, as well as for multiple-user smarthome assistants (operating on a per-person basis) which are not yet available. Speaker recognition, i.e., autonomously recognising a person from their voice, is a well-explored topic in the state-of-the-art within the bounds of data availability, which causes difficulty in real-world use. It is unrealistic to expect a user to willingly provide many minutes or hours of speech data to a device unless it is allowed to constantly record daily life, something which is a modern cause for concern with the virtual home assistant. In this work, we show that data scarcity in speaker recognition can be overcome by collecting only several short spoken sentences of audio from a user and then using extracted Mel-Frequency Cepstral Coefficients (MFCC) data in both supervised and unsupervised learning paradigms to generate synthetic speech, which is then used in a process of transfer learning to better recognise the speaker in question.

Autonomous speaker classification can suffer issues of data scarcity since the user is compared to a large database of many speakers. The most obvious solution to this is to collect more data, but with Smart Home Assistants existing within private environments and potentially listening to private data, this produces an obvious problem of privacy and security~\cite{logsdon2018alexa,dunin2020alexa}. Not collecting more data on the other hand, presents an issue of a large class imbalance between the speaker to classify against the examples of other speakers, producing lower accuracies and less trustworthy results~\cite{babbar2019data}, which must be solved for purposes such as biometrics since results must be trusted when used for security. In this study, weighting of errors is performed to introduce balance, but it is noted that the results still have room for improvement regardless.

Data augmentation is the idea that useful new data can be generated by algorithms or models that would improve the classification of the original, scarce dataset. A simple but prominent example of this is the warping, flipping, mirroring and noising of images to better prepare image classification algorithms~\cite{wang2017effectiveness}. A more complex example through generative models can be seen in recent work that utilise methods such as the Generative Adversarial Network (GAN) to create synthetic data which itself also holds useful information for learning from and classification of data~\cite{zhu2018emotion,frid2018synthetic}. Although image classification is the most common and most obvious application of generative models for data augmentation, recent works have also enjoyed success in augmenting audio data for sound classification~\cite{yang2018se,madhu2019data}. This work extends upon a previous conference paper that explored the hypothesis \textit{``can the synthetic data produced by a generative model aid in speaker recognition?"}~\cite{bird2020overcoming} within which a Character-Level Recurrent Neural Network (RNN), as a generative model, produced synthetic data useful for learning from in order to recognise the speaker. This work extends upon these preliminary experiments by the following:
\begin{enumerate}
    \item The extension of the dataset to more subjects from multiple international backgrounds and the extraction of the MFCCs of each subject;
    \item Benchmarking of a Long Short Term Memory (LSTM) architecture for 64, 128, 256 and 512 LSTM units in one to three hidden layers towards reduction of loss in generating synthetic data. The best model is selected as the candidate for the LSTM data generator;
    \item The inclusion of OpenAI's GPT-2 model as a data generator in order to compare the approaches of supervised (LSTM) and attention-based (GPT-2) methods for synthetic data augmentation for speaker classification.
\end{enumerate}
The scientific contributions of this work, thus, are related to the application of synthetic MFCCs for improvement of speaker recognition. A diagram of the experiments can be observed in Figure \ref{fig:smalldiagram}. To the authors' knowledge, the paper that this work extends is the first instance of this research being explored\footnote{to the best of our knowledge and based on literature review.}. The best LSTM and the GPT-2 model are tasked with generating 2,500, 5,000, 7,500, and 10,000 synthetic data objects for each subject after learning from the scarce datasets extracted from their speech. A network then learns from these data and transfers their weights to a network aiming to learn and classify the real data, and many show an improvement. For all subjects, the results show that several of the networks perform best after experiencing exposure to synthetic data. 

The remainder of this article is as follows. Section \ref{sec1:background} initially explores important scientific concepts of the processes followed by this work and also the current State-of-the-Art in the synthetic data augmentation field. Following this, Section \ref{sec2:method} then outlines the method followed by this work including data collection, synthetic data augmentation, MFCC extraction to transform audio into a numerical dataset, and finally the learning processes followed to achieve results. The final results of the experiments are then discussed in Section \ref{sec:results}, and then future work is outlined and conclusions presented in Section \ref{sec4:futureworkconclusion}.

\section{Background and Related Work}
\label{sec1:background}


Verification of a speaker is the process of identifying a single individual against many others by spoken audio data~\cite{poddar2017speaker}. That is, the recognition of a set of the person's speech data $X$ specifically from a speech set $Y$ where $X \in Y$. In the simplest sense, this can be given as a binary classification problem; for each data object $o$, is $o \in X$? Is the speaker to be recognised speaking, or is it another individual? Speaker recognition is important for social HRI~\cite{mumolo2003distant} (the robot's perception of the individual based on their acoustic utterances), Biometrics~\cite{ratha2001automated}, and Forensics~\cite{rose2002forensic} among many others. In~\cite{hasan2004speaker}, researchers found relative ease of classifying 21 speakers from a limited set, but the problem becomes more difficult as it becomes more realistic, where classifying a speaker based on their utterances is increasingly difficult as the dataset grows~\cite{nagrani2017voxceleb,yadav2018learning,zeinali2019but}. In this work, the speaker is recognised from many thousands of other examples of human speech from the Flickr8k speakers dataset. 

\subsection{LSTM and GPT-2}
Long Short Term Memory (LSTM) is a form of Artificial Neural Network in which multiple RNNs will learn from previous states as well as the current state. Initially, the LSTM selects data to delete: 
\begin{equation}
    \mathop f\nolimits_{\text{t}} = \sigma \left( {\mathop W\nolimits_{f} \cdot \left[ {\mathop h\nolimits_{t - 1} ,\mathop x\nolimits_{t} } \right] + \mathop b\nolimits_{f} } \right),
\end{equation}
\noindent where $W_f$ are the weights of the units, $h_{t=1}$ is the output at $t=1$, $x_t$ are inputs and $b_f$ is an applied bias. Data to be stored is then selected based on input $i$, generating $C_t$ values:
\begin{equation}
\mathop o\nolimits_{\text{t}} = \sigma \left( {\mathop W\nolimits_{i} \cdot \left[ {\mathop h\nolimits_{t - 1} ,\mathop x\nolimits_{t} } \right] + \mathop b\nolimits_{i} } \right),
\end{equation}

\begin{equation}
\mathop {\hat{C}}\nolimits_{\text{t}} = \tanh \left( {\mathop W\nolimits_{c} \cdot \left[ {\mathop h\nolimits_{t - 1} ,\mathop x\nolimits_{t} } \right] + \mathop b\nolimits_{c} } \right).
\end{equation}

A convolutional operation updates values:
\begin{equation}
C_{t}  =  f_{t} * C_{t-1} + i_{t} *  \tilde{C}_{t} .
\end{equation}

Output $o_t$ is presented, and the hidden state is updated:
\begin{equation}
\mathop o\nolimits_{\text{t}} = \sigma \left( {\mathop W\nolimits_{o} \cdot \left[ {\mathop h\nolimits_{t - 1} ,\mathop x\nolimits_{t} } \right] + \mathop b\nolimits_{o} } \right),
\end{equation}
\begin{equation}
h_{t} = o_{t} * tanh(C_{t}).
\end{equation}
Due to the observed consideration of previous data, it is often found that time dependent data are very effectively classified due to the memory-like nature of the LSTM. LSTM is thus a particularly powerful technique in terms of speech recognition~\cite{graves2013hybrid} due to the temporal nature of the data~\cite{belin2000voice}. \\

In addition to LSTM, this study also considers OpenAI's Generative Pretrained Transformer 2 (GPT-2) model~\cite{radford2018improving,radford2019language} as a candidate for producing synthetic MFCC data to improve speaker recognition. The model in question, \textit{335M}, is much deeper than the LSTMs explored at 12 layers. In the OpenAI paper, the GPT modelling is given for a set of samples $x_1, x_2, ..., x_n$ composed of variable symbol-sequence lengths $s_1, s_2, ..., s_n$ factorised by joint probabilities over symbols as the product of conditional probabilities~\cite{jelinek1980interpolated,bengio2003neural}:
\begin{equation}
    p(x) =  \prod_{i=1}^n p(s_{i}  | s_{1}, ..., s_{n-1}).
\end{equation}
Attention is given to various parts of the input vector:
\begin{equation}
    Attention(Q,K,V) = softmax \left(  \frac{QK^T}{  \sqrt{d_{k}}  }  \right) V,
\end{equation}
where Q is the \textit{query} i.e., the single object in the sequence, in this case, a word. K are the keys, which are vector representations of the input sequence, and V are values as vector representations of all words in the sequence. In the initial encoder, decoder, and attention blocks $Q=V$ whereas later on the attention block that takes these outputs as input, $Q \neq V$ since both are derived from the block's 'memory'. In order to combine multiple queries, that is, to consider previously learnt rules from text, multi-headed attention is presented:
\begin{equation}
\begin{aligned}
    MultiHead(Q,K,V) = Concatenate(head_{1}, ..., head_{h})W^{O} \\
    head_{i} = Attention(QW^{Q}_{i}, KW^{K}_{i}, VW^{V}_{i}).
\end{aligned}
\end{equation}
As in the previous equation, it can be seen that previously learnt $h$ projections $d_Q, d_K$ and $d_V$ are also considered given that the block has multiple heads. The above equations and further detail on both attention and multi-headed attention can be found in \cite{vaswani2017attention}. The unsupervised nature of the GPT models is apparent since $Q, K$ and $V$ are from the same source. 

Importantly, GPT produces output with consideration not only to input, but also to the task. The GPT-2 model has been shown to be a powerful state-of-the-art tool for language learning, understanding, and generation; researchers noted that the model could be used with ease to generate realistic propaganda for extremist terrorist groups~\cite{solaiman2019release}, as well as noting that generated text by the model was difficult to detect~\cite{gehrmann2019gltr,wolff2020attacking,adelani2020generating}. The latter aforementioned papers are promising, since a difficulty of detection suggests statistical similarities, which are likely to aid in the problem of improving classification accuracy of a model by exposing it to synthetic data objects output by such a model. 

\subsection{Dataset Augmentation through Synthesis}
\label{subsec:dataaugmentation}

Similarities between real-life experiences and imagined perception have shown in psychological studies that the human imagination, though mentally augmenting and changing situations~\cite{beres1960perception}, aids in improving the human learning process~\cite{egan1989memory,heath2008exploring,macintyre2012emotions,egan2014imagination}. The importance of this ability in the learning process shows the usefulness of data augmentation in human learning, and as such, is being explored as a potential solution to data scarcity and quality in the machine learning field. Even though the synthetic data may not be realistic alone, minute similarities between it and reality allow for better pattern recognition. 

The idea of data augmentation as the first stage in fine-tune learning is inspired by the aforementioned findings, and follows a similar approach. Synthetic data is generated by learning from the real data, and algorithms are exposed to them in a learning process prior to the learning process of real data; this is then compared to the classical approach of learning from the data solely, where the performance of the former model compared to the latter shows the effect of the data augmentation learning process. Much of the work is recent, many from the last decade, and a pattern of success is noticeable for many prominent works when comparing the approach to the classical method of learning from real data alone. 

As described, the field of exploring augmented data to improve classification algorithms is relatively young, but there exist several prominent works that show success in applying this approach. When augmented data from the SemEval dataset is learned from by a Recurrent Neural Network (RNN), researchers found that the overall best F-1 score was achieved for relation classification in comparison to the model only learning from the dataset itself~\cite{xu2016improved}. Due to data scarcity in the medical field, classification of liver lesions~\cite{frid2018synthetic} and Alzheimer's Disease~\cite{shin2018medical} have also shown improvement when the learning models (CNNs) considered data augmented by Convolutional GANs. In Natural Language Processing, it was found that word augmentation aids to improve sentence classification by both CNN and RNN models~\cite{kobayashi2018contextual}. The \textit{DADA} model has been suggested as a strong method to produce synthetic data for improvement of data-scarce problems through the Deep Adversarial method via a dedicated discriminator network aiming to augment data specifically~\cite{zhang2019dada} which has noted success in machine learning problems~\cite{barz2020deep}. 

Data augmentation has shown promise in improving multi-modal emotion recognition when considering audio and images~\cite{huang2018multimodal}, digital signal classification~\cite{tang2018digital}, as well as for a variety of audio classification problems such as segments of the Hub500 problem~\cite{park2019specaugment}. Additionally, synthetic data augmentation of mel-spectrograms have shown to improve acoustic scene recognition~\cite{yang2018se}. Realistic text-to-speech is achieved by producing realistic sound such as the Tacotron model~\cite{wang2017tacotron} when considering the textual representation of the audio being considered, and reverse engineering the model to produce audio based on text input. A recent preliminary study showed GANs may be able to aid in producing synthetic data for speaker recognition~\cite{chien2018adversarial}. 

The temporal models considered in this work to generate synthetic speech data have recently shown success in generating acoustic sounds~\cite{eck2002finding} and accurate timeseries data~\cite{senjyu2006application}, written text~\cite{pawade2018story,sha2018order}, artistic images~\cite{gregor2015draw}. Specifically, temporal models are also observed to be successful in generating MFCC data~\cite{wang2017rnn}, which is the data type considered in this work. Many prominent works in speech recognition consider temporal learning to be highly important~\cite{fernandez2007application,he2019streaming,sak2014long} and for generation of likewise temporal data~\cite{valentini2016investigating,wang2017rnn,tachibana2018efficiently} (that is, which this study aims to perform). If it is possible to generate data that bares similarity to the real data, then it could improve the models while also reducing the need for large amounts of real data to be collected.

\section{Proposed Approach}
\label{sec2:method}

\begin{figure*}
    \centering
    \includegraphics[scale=0.9]{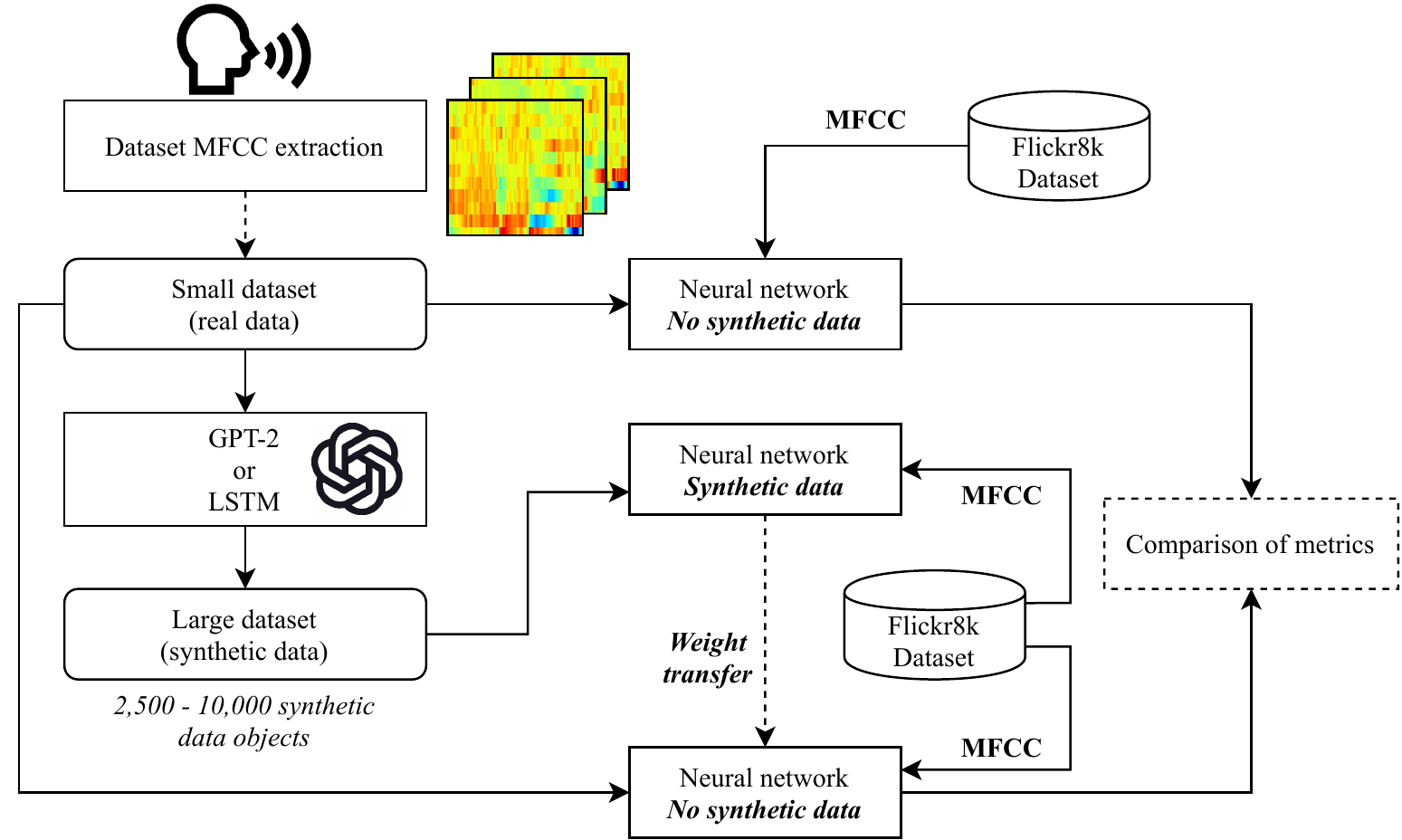}
    \caption{A diagram of the experimental method in this work. Note that the two networks being directly compared are classifying the same data, with the difference being the initial weight distribution either from standard random distribution or transfer learning from GPT-2 and LSTM produced synthetic data.}
    \label{fig:method}
\end{figure*}

This section describes the development of the proposed approach, which can be observed overall in Figure \ref{fig:method}. For each test, five networks are trained in order to gain results. Firstly, a network simply to perform the speaker classification experimen without transfer learning (from a standard random weight distribution). Firstly, a network simply to perform the speaker classification experiment. Produced by LSTM and GPT-2, synthetic data is used to train another network, of which the weights are used to train a final network as an initial distribution to perform the same experiment as described in the first network (classifying the speaker's real data from Flickr8k speakers). Thus, the two networks leading to the final classification score in the diagram are directly comparable since they are learning from the same data, they differ only in initial weight distribution (where the latter network has weights learnt from synthetic data).

\subsection{Real and Synthetic Data Collection}
The data collected, as previously described, presents a binary classification problem. That is, whether or not the individual in question is the one producing the acoustic utterance. 

The large corpus of data for the \textit{``not the speaker"} class is gathered via the Flickr8k dataset~\cite{harwath2015deep} which contains 40,000 individual utterances describing 8,000 images by a large number of speakers which is unspecified by the authors. MFCCs are extracted (described in Section \ref{subsec:mfcc}) to generate temporal numerical vectors which represent a short amount of time from each audio clip. 100,000 data objects are selected through 50 blocks of 1,000 objects and then 50,000 other data objects selected randomly from the remainder of the dataset. This is performed so the dataset contains individual's speech at length as well as short samples of many other thousands of speakers also. 

\begin{table*}[]
\footnotesize
\centering
\caption{Information regarding the data collection from the seven subjects}
\label{tab:subjects}
\begin{tabular}{@{}lllllll@{}}
\toprule
\textbf{Subject}           & \textbf{Sex} & \textbf{Age} & \textbf{Nationality} & \textbf{Dialect} & \textbf{Time Taken (s)} & \textbf{\begin{tabular}[c]{@{}l@{}}Real\\Data\end{tabular}} \\ \midrule
\textit{\textbf{1}}        & M            & 23           & British              & Birmingham              & 24                      & 4978                           \\
\textit{\textbf{2}}        & M            & 24           & American             & Florida          & 13                      & 2421                           \\
\textit{\textbf{3}}        & F            & 28           & Irish                & Dublin                  & 12                      & 2542                           \\ 
\textit{\textbf{4}}        & F            & 30           & British                & London                  & 12                      & 2590                           \\ 
\textit{\textbf{5}}        & F            & 40           & British                & London                  & 10                      & 2189                           \\ 
\textit{\textbf{6}}        & M            & 21           & French                & Paris                  & 8                      & 1706                           \\ 
\textit{\textbf{7}}        & F            & 23           & French                & Paris                  & 9                      & 1952                           \\ 
\midrule
\textit{\textbf{Flickr8K}} &              &              &                      &                         &                         & 100000                       \\ \bottomrule
\end{tabular}
\end{table*}

To gather data for recognising speakers, seven subjects are considered. Information on the subjects can be seen in Table \ref{tab:subjects}. Subjects speak five random Harvard Sentences sentences from the \textit{IEEE recommended practice for speech quality measurements}~\cite{rothauser1969ieee}, and so contain most of the spoken phonetic sounds in the English language~\cite{bird2019phoneme2}. Importantly, this is a user-friendly process, because it requires only a few short seconds of audio data. The longest time taken was by subject 1 in 24 seconds producing 4978 data objects and the shortest were the two French individuals who required 8 and 9 seconds respectively to speak the five sentences. All of the audio data were recorded using consumer-available recording devices such as smartphones and computer headsets. Synthetic datasets are generated following the learning processes of the best LSTM and the GPT-2 model, where the probability of the next character is decided upon depending on the learning algorithm and are generated in blocks of 1,000 within a loop and the final line is removed (since it was often within the cutoff point of the 1,000-character block). Illogical lines of data (those that did not have 26 comma separated values and a class) were removed, but were observed to be somewhat rare as both the LSTM and GPT-2 models had learnt the data format relatively well since it was uniform throughout. The format, throughout the datasets, was a uniform 27 comma separated values where the values were all numerical and the final value was `1' followed by a line break character.

\subsection{Feature Extraction}
\label{subsec:mfcc}

The nature of acoustic data is that the previous and following points of data from a single point in particular are also related to the class. Audio data is temporal in this regard, and thus classification of a single point in time is an extremely difficult problem~\cite{xiong2003comparing,bird2019phoneme}. In order to overcome this issue features are extracted from the wave through a sliding window approach. The statistical features extracted in this work are the first 26 Mel-Frequency Cepstral Coefficients due to findings in the scientific state of the art arguing for their prominence over other methods~\cite{muda2010voice,sahidullah2012design}. Sliding windows are set to a length of 0.025 seconds with a step of 0.01 seconds and extraction is performed as follows:
\begin{enumerate}
\item The Fourier Transform of time window data $\omega$ is calculated:
\begin{equation}
X(j\omega)=\int_{-\infty}^\infty x(t) e^{-j\omega t} dt.
\end{equation}
\item The powers from the FT are mapped to the Mel-scale, that is, the human psychological scale of audible pitch~\cite{stevens1937scale} via a triangular temporal window. 
\item The power spectrum is considered and $log$s of each of their powers are taken.
\item The derived Mel-log powers are then treated as a signal, and a Discrete Cosine Transform (DCT) is calculated:
\begin{equation}
\begin{aligned}
X_k =  \sum_{n=0}^{N-1} x_n  cos  \begin{bmatrix}  \frac{\pi}{N} (n+ \frac{1}{2}) k \end{bmatrix} \\     
k=0,...,N-1,
\end{aligned}
\end{equation}
\noindent where $x$ is the array of length $N$, $k$ is the index of the output coefficient being calculated, where $N$ real numbers $x_{0} ... x_{n-1}$ are transformed into the $N$ real numbers $X_{0} ... X_{n-1}$. 
\end{enumerate}
The amplitudes of the spectrum produced are taken as the \textit{MFCCs}. The resultant data then provides a mathematical description of wave behaviour in terms of sounds, each data object made of 26 attributes produced from the sliding window are then treated as the input attributes for the neural networks for both speaker recognition and also synthetic data generation (with a class label also). 

This process is performed for all of the selected Flickr8K data as well as the real data recorded from the subjects. The MFCC data from each of the 7 subjects' audio recordings is used as input to the LSTM and GPT-2 generative models for training and subsequent data augmentation.

\subsection{Speaker Classification Learning Process}
For each subject, the Flickr data and recorded audio forms the basis dataset and the speaker recognition problem. Eight datasets for transfer learning are then formed on a per-subject basis, which are the aforementioned data plus $2500, 5000, 7500$ and $10000$ synthetic data objects generated by either the LSTM or the GPT-2 models. LSTM has a standard dropout of 0.2 between each layer.

The baseline accuracy for comparison is given as ``Synth. Data: 0" later in Table \ref{tab:bigresults} which denotes a model that has not been exposed to any of the synthetic data. This baseline gives scores that are directly comparable to identical networks with their initial weight distributions being those trained to classify synthetic data generated for the subject, which is then used to learn from the real data. As previously described, the two sets of synthetic data to expose the models to during pre-training of the real speaker classification problem are generated by either an LSTM or a GPT-2 language model. Please note that due to this, the results presented have no baring on whether or not the network could classify the synthetic data well or otherwise, the weights are simply used as the initial distribution for the same problem. If the pattern holds that the transfer learning networks achieve better results than the networks which have not been trained on such data, it argues the hypothesis that speaker classification can be improved when considering either of these methods of data augmentation. This process can be observed in Figure \ref{fig:method}. 

For the Deep Neural Network that classifies the speaker, a topology of three hidden layers consisting of 30, 7, and 29 neurons respectively with ReLu activation functions and an ADAM optimiser~\cite{kingma2014adam} is initialised. These hyperparameters are chosen due to a previous study that performed a genetic search of neural network topologies for the classification of phonetic sounds in the form of MFCCs~\cite{bird2020optimisation}. The networks are given an unlimited number of epochs to train, only ceasing through a set early stopping callback of 25 epochs with no improvement of ability. The best weights are restored before final scores are calculated. This is allowed in order to make sure that all models stabilise to an asymptote and reduce the risk of stopping models prior to them achieving their potential best abilities. 

Classification errors are weighted equally by class prominence since there exists a large imbalance between the speaker and the rest of the data. All of the LSTM experiments performed in this work were executed on an Nvidia GTX980Ti GPU, while the GPT-2 experiment was performed on an Nvidia Tesla K80 GPU provided by Google Colab. 

\section{Results}
\label{sec:results}

\begin{figure}
    \centering
    \includegraphics[scale=0.5]{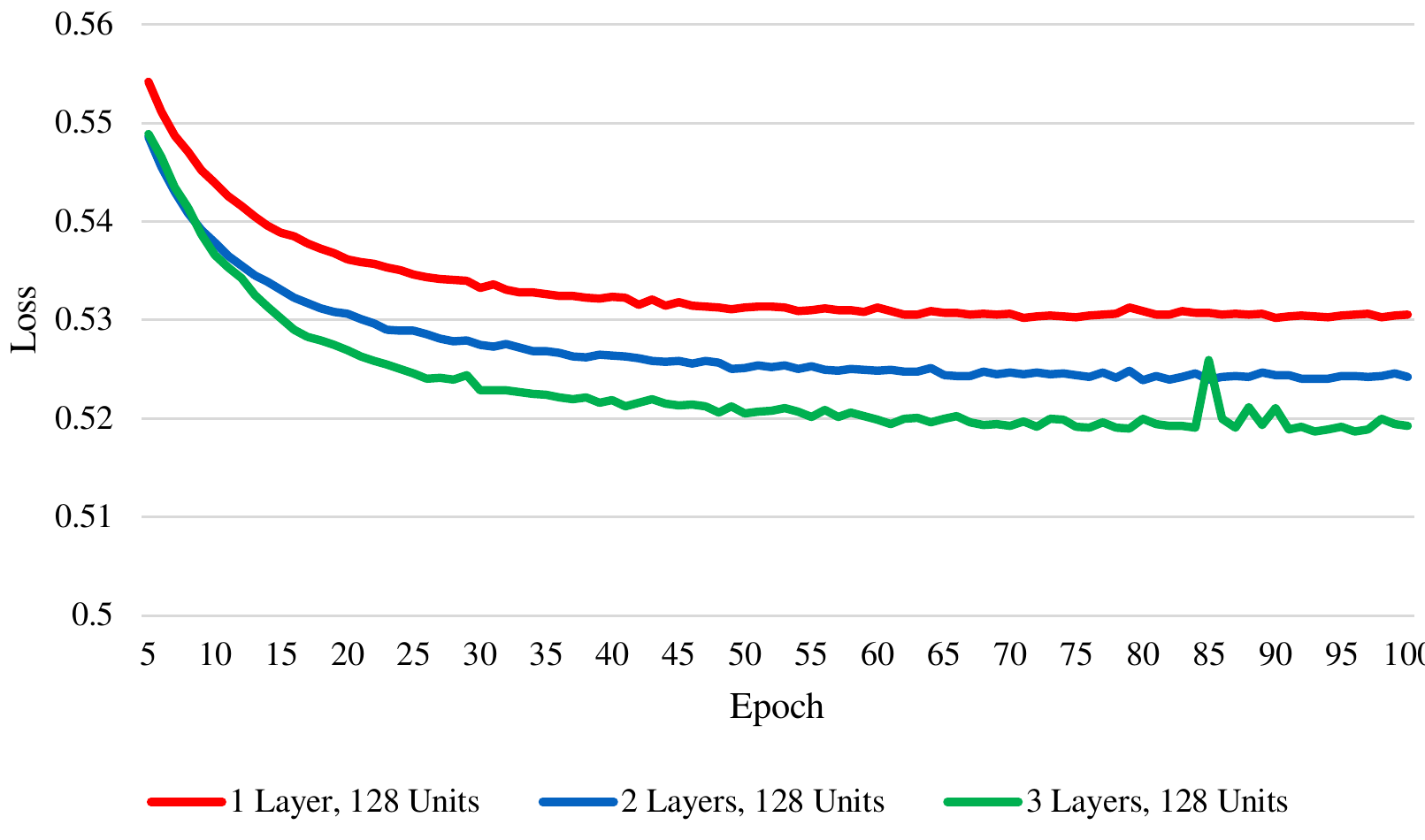}
    \caption{The training processes of the best performing models in terms of loss, separated for readability purposes. Results are given for a benchmarking experiment on all of the dataset rather than an individual.}
    \label{fig:128-lstms-benchmark}
\end{figure}

\begin{figure}
    \centering
    \includegraphics[scale=0.5]{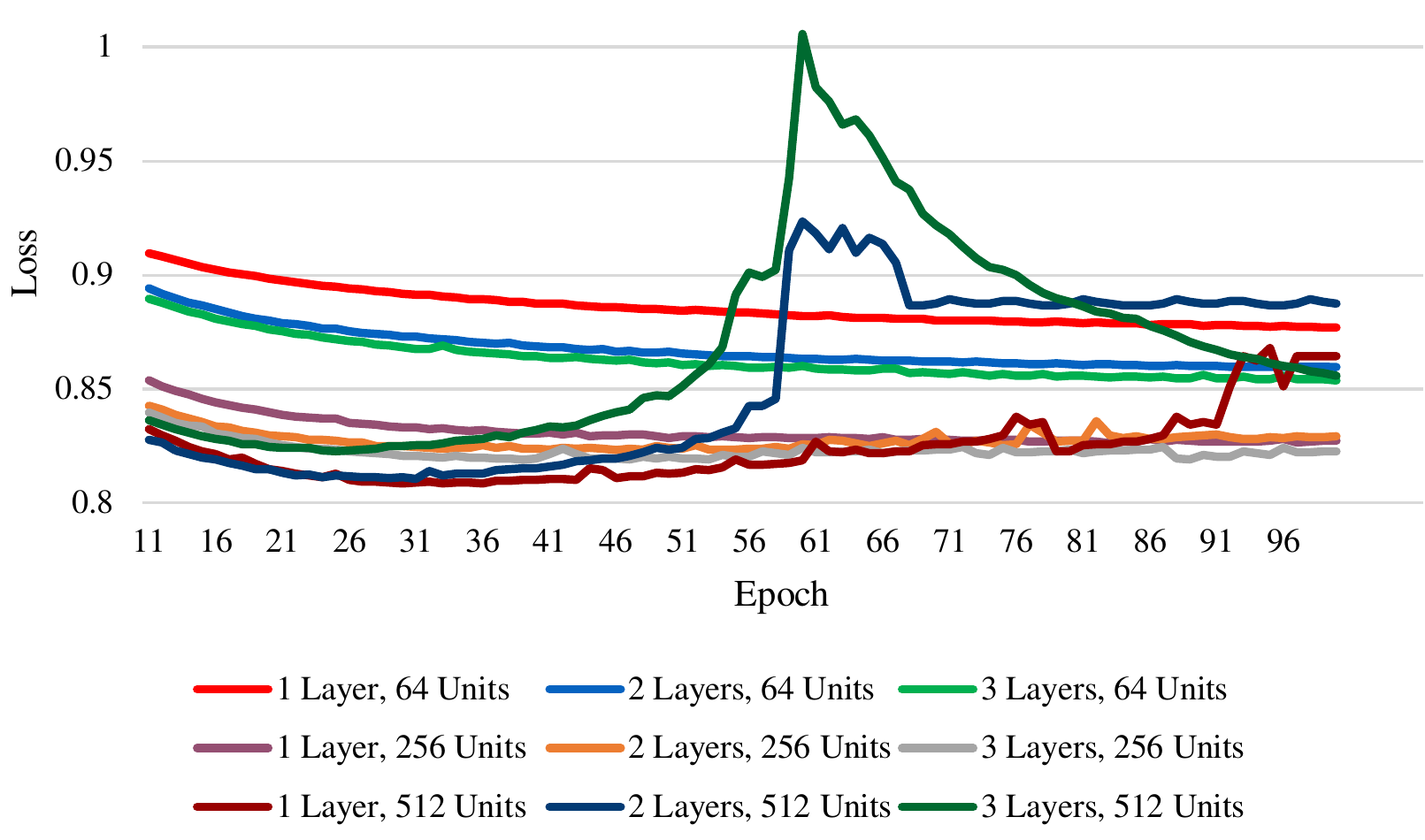}
    \caption{The training processes of LSTMs with 64, 256, and 512 units in 1-3 hidden layers, separated for readability purposes. Results are given for a benchmarking experiment on all of the dataset rather than an individual.}
    \label{fig:bad-lstms-benchmark}
\end{figure}

\begin{figure}
    \centering
    \includegraphics[scale=0.5]{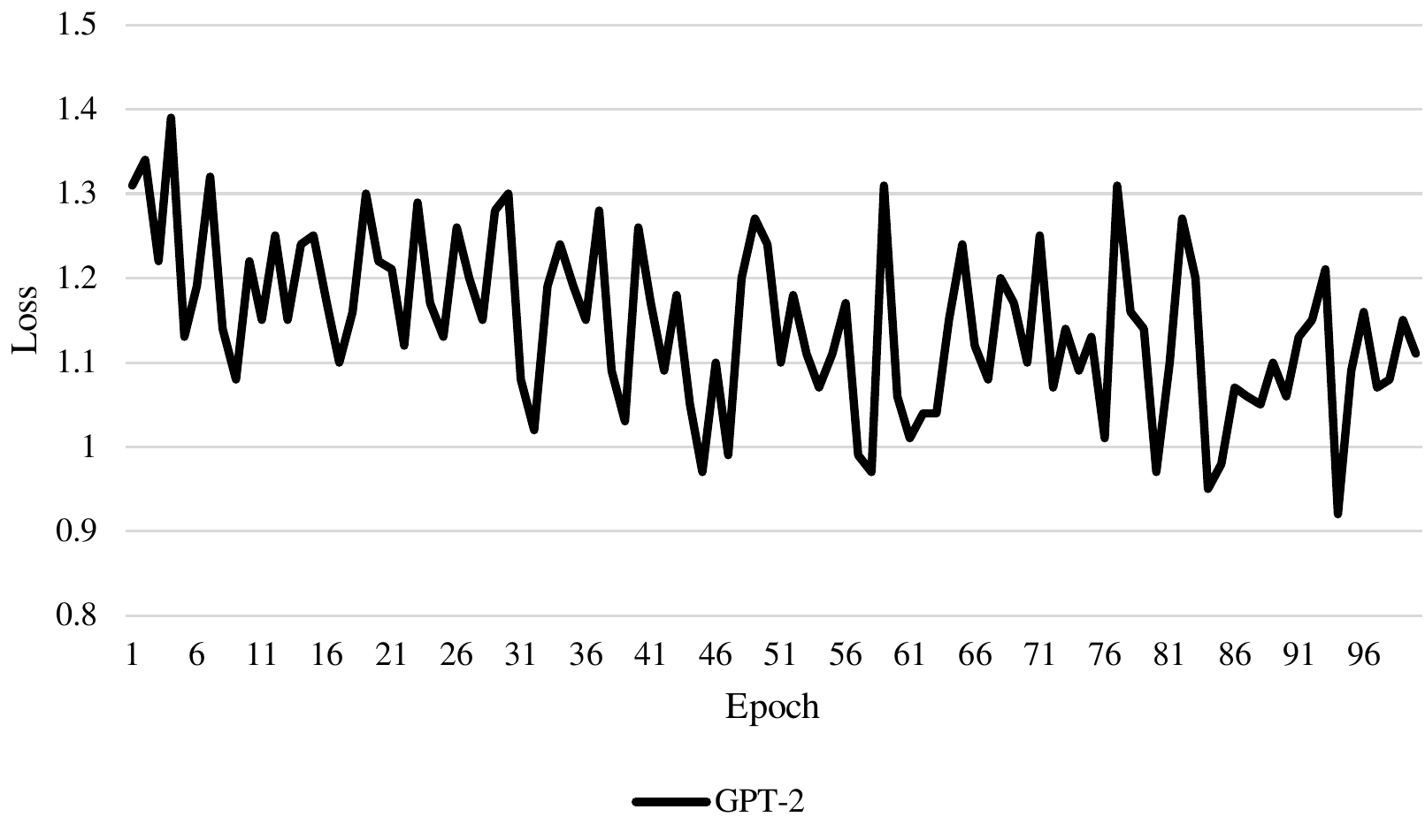}
    \caption{The training process of the GPT-2 model. Results are given for a benchmarking experiment on all of the dataset rather than an individual.}
    \label{fig:gpt-benchmark}
\end{figure}

\begin{table}[]
\centering
\caption{Best epochs and their losses for the 12 LSTM Benchmarks and GPT-2 training process. All models are benchmarked on the whole set of subjects for 100 epochs each in order to search for promising hyperparameters.}
\label{tab:benchmarking-models}
\begin{tabular}{@{}lll@{}}
\toprule
\textbf{Model}    & \textbf{Best Loss} & \textbf{Epoch} \\ \midrule
LSTM(64)          & 0.88           & 99             \\
LSTM(64,64)       & 0.86           & 99             \\
LSTM(64,64,64)    & 0.85           & 99             \\
LSTM(128)         & 0.53  & 71             \\
LSTM(128,128)     & 0.53            & 80             \\
LSTM(128,128,128) & \textbf{0.52}            & 93             \\
LSTM(256)         & 0.83           & 83             \\
LSTM(256,256)     & 0.82            & 46             \\
LSTM(256,256,256) & 0.82            & 39             \\
LSTM(512)         & 0.81           & 33             \\
LSTM(512,512)     & 0.81           & 31             \\
LSTM(512,512,512) & 0.82           & 25             \\
GPT-2             & 0.92               & 94             \\ \bottomrule
\end{tabular}
\end{table}

Table \ref{tab:benchmarking-models} shows the best results discovered for each LSTM hyperparameter set and the GPT-2 model. Figures \ref{fig:128-lstms-benchmark} and \ref{fig:bad-lstms-benchmark} show the epoch-loss training processes for the LSTMs separated for readability purposes and Figure \ref{fig:gpt-benchmark} shows the same training process for the GPT-2 model. These generalised experiments for all data provide a tuning point for synthetic data to be generated for each of the individuals (given respective personally trained models). LSTMs with 128 hidden units far outperformed the other models, which were also sometimes erratic in terms of their attempt at loss reduction over time. The GPT-2 model is observed to be especially erratic, which is possibly due to its unsupervised attention-based approach. 

Although some training processes were not as smooth as others, manual exploration showed that acceptable sets of data were able to be produced.

\subsection{Transfer Learning for Data-scarce Speaker Recognition}

\begin{table*}[]
\centering
\caption{Results of the experiments for all subjects. Best models for each Transfer Learning experiment are bold, and the best overall result per-subject is also underlined. Red font denotes a synthetic data-exposed model that scored lower than the classical learning approach. }
\label{tab:bigresults}
\footnotesize
\begin{tabular}{@{}llllllllll@{}}
\toprule
\multicolumn{1}{c}{}                                   &                                                                                  & \multicolumn{4}{l}{\textbf{LSTM}}                                                                                                           & \multicolumn{4}{l}{\textbf{GPT-2}}                                                                                                                         \\ \cmidrule(l){3-10} 
\multicolumn{1}{c}{\multirow{-2}{*}{\textbf{Subject}}} & \multirow{-2}{*}{\textbf{\begin{tabular}[c]{@{}l@{}}Synth.\\ Data\end{tabular}}} & \textit{\textbf{Acc.}}       & \textit{\textbf{F1}}        & \textit{\textbf{Prec.}}     & \multicolumn{1}{l|}{\textit{\textbf{Rec.}}}      & \textit{\textbf{Acc.}}                & \textit{\textbf{F1}}                 & \textit{\textbf{Prec.}}              & \textit{\textbf{Rec.}}               \\ \midrule
                                                       & \textit{\textbf{0}}                                                              & 93.57                        & 0.94                        & 0.93                        & \multicolumn{1}{l|}{0.93}                        & 93.57                                 & 0.94                                 & 0.93                                 & 0.93                                 \\
                                                       & \textit{\textbf{2500}}                                                           & {\ul \textbf{99.5}}          & {\ul \textbf{$\sim$1}}      & {\ul \textbf{$\sim$1}}      & \multicolumn{1}{l|}{{\ul \textbf{$\sim$1}}}      & 97.32                                 & 0.97                                 & 0.97                                 & 0.97                                 \\
                                                       & \textit{\textbf{5000}}                                                           & 97.37                        & 0.97                        & 0.97                        & \multicolumn{1}{l|}{0.97}                        & \textbf{97.77}                        & \textbf{0.98}                        & \textbf{0.98}                        & \textbf{0.98}                        \\
                                                       & \textit{\textbf{7500}}                                                           & \textbf{99.33}               & \textbf{0.99}               & \textbf{0.99}               & \multicolumn{1}{l|}{\textbf{0.99}}               & 99.2                                  & 0.99                                 & 0.99                                 & 0.99                                 \\
\multirow{-5}{*}{\textbf{1}}                           & \textit{\textbf{10000}}                                                          & 99.1                         & 0.99                        & 0.99                        & \multicolumn{1}{l|}{0.99}                        & \textbf{99.3}                         & \textbf{0.99}                        & \textbf{0.99}                        & \textbf{0.99}                        \\ \cmidrule(r){1-1}
                                                       & \textit{\textbf{0}}                                                              & 95.13                        & 0.95                        & 0.95                        & \multicolumn{1}{l|}{0.95}                        & 95.13                                 & 0.95                                 & 0.95                                 & 0.95                                 \\
                                                       & \textit{\textbf{2500}}                                                           & \textbf{99.6}                & \textbf{$\sim$1}            & \textbf{$\sim$1}            & \multicolumn{1}{l|}{\textbf{$\sim$1}}            & 99.5                                  & $\sim$1                              & $\sim$1                              & $\sim$1                              \\
                                                       & \textit{\textbf{5000}}                                                           & \textbf{99.5}                & \textbf{$\sim$1}            & \textbf{$\sim$1}            & \multicolumn{1}{l|}{\textbf{$\sim$1}}            & 99.41                                 & 0.99                                 & 0.99                                 & 0.99                                 \\
                                                       & \textit{\textbf{7500}}                                                           & {\ul \textbf{99.7}}          & {\ul \textbf{$\sim$1}}      & {\ul \textbf{$\sim$1}}      & \multicolumn{1}{l|}{{\ul \textbf{$\sim$1}}}      & {\ul \textbf{99.7}}                   & {\ul \textbf{$\sim$1}}               & {\ul \textbf{$\sim$1}}               & {\ul \textbf{$\sim$1}}               \\
\multirow{-5}{*}{\textbf{2}}                           & \textit{\textbf{10000}}                                                          & \textbf{99.42}               & \textbf{0.99}               & \textbf{0.99}               & \multicolumn{1}{l|}{\textbf{0.99}}               & 99.38                                 & 0.99                                 & 0.99                                 & 0.99                                 \\ \cmidrule(r){1-1}
                                                       & \textit{\textbf{0}}                                                              & 96.58                        & 0.97                        & 0.97                        & \multicolumn{1}{l|}{0.97}                        & 96.58                                 & 0.97                                 & 0.97                                 & 0.97                                 \\
                                                       & \textit{\textbf{2500}}                                                           & \textbf{99.2}                & \textbf{0.99}               & \textbf{0.99}               & \multicolumn{1}{l|}{\textbf{0.99}}               & 98.41                                 & 0.98                                 & 0.98                                 & 0.98                                 \\
                                                       & \textit{\textbf{5000}}                                                           & 98.4                         & 0.98                        & 0.98                        & \multicolumn{1}{l|}{0.98}                        & \textbf{99}                           & \textbf{0.99}                        & \textbf{0.99}                        & \textbf{0.99}                        \\
                                                       & \textit{\textbf{7500}}                                                           & \textbf{99.07}               & \textbf{0.99}               & \textbf{0.99}               & \multicolumn{1}{l|}{\textbf{0.99}}               & 98.84                                 & 0.99                                 & 0.99                                 & 0.99                                 \\
\multirow{-5}{*}{\textbf{3}}                           & \textit{\textbf{10000}}                                                          & 98.44                        & 0.98                        & 0.98                        & \multicolumn{1}{l|}{0.98}                        & {\ul \textbf{99.47}}                  & {\ul \textbf{0.99}}                  & {\ul \textbf{0.99}}                  & {\ul \textbf{0.99}}                  \\ \cmidrule(r){1-1}
                                                       & \textit{\textbf{0}}                                                              & 98.5                         & 0.99                        & 0.99                        & \multicolumn{1}{l|}{0.99}                        & 98.5                                  & 0.99                                 & 0.99                                 & 0.99                                 \\
                                                       & \textit{\textbf{2500}}                                                           & {\color[HTML]{FE0000} 97.86} & {\color[HTML]{FE0000} 0.98} & {\color[HTML]{FE0000} 0.98} & \multicolumn{1}{l|}{{\color[HTML]{FE0000} 0.98}} & \textbf{99.42}                        & \textbf{0.99}                        & \textbf{0.99}                        & \textbf{0.99}                        \\
                                                       & \textit{\textbf{5000}}                                                           & \textbf{99.22}               & \textbf{0.99}               & \textbf{0.99}               & \multicolumn{1}{l|}{\textbf{0.99}}               & {\color[HTML]{FE0000} 97.75}          & {\color[HTML]{FE0000} 0.98}          & {\color[HTML]{FE0000} 0.98}          & {\color[HTML]{FE0000} 0.98}          \\
                                                       & \textit{\textbf{7500}}                                                           & {\color[HTML]{FE0000} 97.6}  & {\color[HTML]{FE0000} 0.98} & {\color[HTML]{FE0000} 0.98} & \multicolumn{1}{l|}{{\color[HTML]{FE0000} 0.98}} & {\color[HTML]{FE0000} \textbf{98.15}} & {\color[HTML]{FE0000} \textbf{0.98}} & {\color[HTML]{FE0000} \textbf{0.98}} & {\color[HTML]{FE0000} \textbf{0.98}} \\
\multirow{-5}{*}{\textbf{4}}                           & \textit{\textbf{10000}}                                                          & 99.22                        & 0.99                        & 0.99                        & \multicolumn{1}{l|}{0.99}                        & {\ul \textbf{99.56}}                  & {\ul \textbf{$\sim$1}}               & {\ul \textbf{$\sim$1}}               & {\ul \textbf{$\sim$1}}               \\ \cmidrule(r){1-1}
                                                       & \textit{\textbf{0}}                                                              & 96.6                         & 0.97                        & 0.97                        & \multicolumn{1}{l|}{0.97}                        & 96.6                                  & 0.97                                 & 0.97                                 & 0.97                                 \\
                                                       & \textit{\textbf{2500}}                                                           & \textbf{99.47}               & \textbf{0.99}               & \textbf{0.99}               & \multicolumn{1}{l|}{\textbf{0.99}}               & 99.23                                 & 0.99                                 & 0.99                                 & 0.99                                 \\
                                                       & \textit{\textbf{5000}}                                                           & 99.4                         & 0.99                        & 0.99                        & \multicolumn{1}{l|}{0.99}                        & \textbf{99.83}                        & \textbf{$\sim$1}                     & \textbf{$\sim$1}                     & \textbf{$\sim$1}                     \\
                                                       & \textit{\textbf{7500}}                                                           & 99.2                         & 0.99                        & 0.99                        & \multicolumn{1}{l|}{0.99}                        & {\ul \textbf{99.85}}                  & {\ul \textbf{$\sim$1}}               & {\ul \textbf{$\sim$1}}               & {\ul \textbf{$\sim$1}}               \\
\multirow{-5}{*}{\textbf{5}}                           & \textit{\textbf{10000}}                                                          & 99.67                        & $\sim$1                     & $\sim$1                     & \multicolumn{1}{l|}{$\sim$1}                     & \textbf{99.78}                        & \textbf{$\sim$1}                     & \textbf{$\sim$1}                     & \textbf{$\sim$1}                     \\ \cmidrule(r){1-1}
                                                       & \textit{\textbf{0}}                                                              & 97.3                         & 0.97                        & 0.97                        & \multicolumn{1}{l|}{0.97}                        & 97.3                                  & 0.97                                 & 0.97                                 & 0.97                                 \\
                                                       & \textit{\textbf{2500}}                                                           & \textbf{99.8}                & \textbf{$\sim$1}            & \textbf{$\sim$1}            & \multicolumn{1}{l|}{\textbf{$\sim$1}}            & 99.75                                 & $\sim$1                              & $\sim$1                              & $\sim$1                              \\
                                                       & \textit{\textbf{5000}}                                                           & 99.75                        & $\sim$1                     & $\sim$1                     & \multicolumn{1}{l|}{$\sim$1}                     & {\color[HTML]{FE0000} 96.1}           & {\color[HTML]{FE0000} 0.96}          & {\color[HTML]{FE0000} 0.96}          & {\color[HTML]{FE0000} 0.96}          \\
                                                       & \textit{\textbf{7500}}                                                           & {\color[HTML]{000000} 97.63} & {\color[HTML]{000000} 0.98} & {\color[HTML]{000000} 0.98} & \multicolumn{1}{l|}{{\color[HTML]{000000} 0.98}} & {\ul \textbf{99.82}}                  & {\ul \textbf{$\sim$1}}               & {\ul \textbf{$\sim$1}}               & {\ul \textbf{$\sim$1}}               \\
\multirow{-5}{*}{\textit{\textbf{6}}}                  & \textit{\textbf{10000}}                                                          & 99.67                        & $\sim$1                     & $\sim$1                     & \multicolumn{1}{l|}{$\sim$1}                     & \textbf{99.73}                        & \textbf{$\sim$1}                     & \textbf{$\sim$1}                     & \textbf{$\sim$1}                     \\ \cmidrule(r){1-1}
                                                       & \textit{\textbf{0}}                                                              & 90.7                         & 0.91                        & 0.91                        & \multicolumn{1}{l|}{0.91}                        & 90.7                                  & 0.91                                 & 0.91                                 & 0.91                                 \\
                                                       & \textit{\textbf{2500}}                                                           & \textbf{99.86}               & \textbf{$\sim$1}            & \textbf{$\sim$1}            & \multicolumn{1}{l|}{\textbf{$\sim$1}}            & 99.78                                 & $\sim$1                              & $\sim$1                              & $\sim$1                              \\
                                                       & \textit{\textbf{5000}}                                                           & \textbf{99.89}               & \textbf{$\sim$1}            & \textbf{$\sim$1}            & \multicolumn{1}{l|}{\textbf{$\sim$1}}            & 99.86                                 & $\sim$1                              & $\sim$1                              & $\sim$1                              \\
                                                       & \textit{\textbf{7500}}                                                           & \textbf{99.91}               & \textbf{$\sim$1}            & \textbf{$\sim$1}            & \multicolumn{1}{l|}{\textbf{$\sim$1}}            & 99.84                                 & $\sim$1                              & $\sim$1                              & $\sim$1                              \\
\multirow{-5}{*}{\textit{\textbf{7}}}                  & \textit{\textbf{10000}}                                                          & {\ul \textbf{99.94}}         & {\ul \textbf{$\sim$1}}      & {\ul \textbf{$\sim$1}}      & \multicolumn{1}{l|}{{\ul \textbf{$\sim$1}}}      & 99.73                                 & $\sim$1                              & $\sim$1                              & $\sim$1                              \\ \midrule
\textit{\textbf{Avg.}}                                 &                                                                                  & 98.43                        & 0.98                        & 0.98                        & 0.98                                             & 98.40                                 & 0.98                                 & 0.98                                 & 0.98                                 \\ \bottomrule
\end{tabular}
\end{table*}

Table \ref{tab:bigresults} shows all of the results for each subject, both with and without exposure to synthetic data. Per-run, the LSTM achieved better results over the GPT-2 in 14 instances whereas the GPT-2 achieved better results over the LSTM in 13 instances. Of the five runs that scored lower than no synthetic data exposure, two were LSTM and three were GPT-2. Otherwise, 51 of the 56 experiments all outperformed the original model without synthetic data exposure and every single subject experienced their best classification result in all cases when the model had been exposed to synthetic data. The best score on a per-subject basis was achieved by exposing the network to data produced by the LSTM three times and the GPT-2 five times (both including Subject 2 where both were best at 99.7\%). The maximum diversion of training accuracy to validation accuracy was $\sim1\%$ showing that although high results were attained, overfitting was relatively low; with more computational resources, k-fold and LOO cross validation are suggested as future works to achieve more accurate measures of variance within classification. 

These results attained show that speaker classification can be improved by exposing the network to synthetic data produced by both supervised and attention-based models and then transferring the weights to the initial problem, which most often scores lower without synthetic data exposure in all cases but five although those subjects still experienced their absolute best result through synthetic data exposure regardless.

\subsection{Comparison to other methods of speaker recognition}
\begin{table}[]
\caption{Comparison of the best models found in this work and other classical methods of speaker recognition (sorted by accuracy)}
\label{tab:bigcomparison}
\footnotesize
\begin{tabular}{@{}clllll@{}}
\toprule
\multicolumn{1}{l}{\textbf{Subject}} & \textbf{Model}                         & \textbf{Acc.}  & \textbf{F-1} & \textbf{Prec.} & \textbf{Rec.} \\ \midrule
\multirow{7}{*}{\textbf{1}}          & \textit{\textbf{DNN (LSTM TL 2500)}}   & \textbf{99.5}  & \textbf{$\sim$1}   & \textbf{$\sim$1}     & \textbf{$\sim$1}    \\
                                     & \textit{\textbf{DNN (GPT-2 TL 5000)}}  & 97.77          & 0.98         & 0.98           & 0.98          \\
                                     & \textit{\textbf{SMO}}                  & 97.71          & 0.98         & 0.95           & 0.95          \\
                                     & \textit{\textbf{Random Forest}}        & 97.48          & 0.97         & 0.97           & 0.97          \\
                                     & \textit{\textbf{Logistic Regression}}  & 97.47          & 0.97         & 0.97           & 0.97          \\
                                     & \textit{\textbf{Bayesian Network}}     & 82.3           & 0.87         & 0.96           & 0.82          \\
                                     & \textit{\textbf{Naive Bayes}}          & 78.96          & 0.84         & 0.953          & 0.77          \\ \cmidrule(r){1-1}
\multirow{7}{*}{\textbf{2}}          & \textit{\textbf{DNN (LSTM TL 7500)}}   & \textbf{99.7}  & \textbf{$\sim$1}   & \textbf{$\sim$1}     & \textbf{$\sim$1}    \\
                                     & \textit{\textbf{DNN (GPT-2 TL 7500)}}  & 99.7           & $\sim$1            & $\sim$1              & $\sim$1             \\
                                     & \textit{\textbf{SMO}}                  & 98.94          & 0.99         & 0.99           & 0.99          \\
                                     & \textit{\textbf{Logistic Regression}}  & 98.33          & 0.98         & 0.98           & 0.98          \\
                                     & \textit{\textbf{Random Forest}}        & 98.28          & 0.98         & 0.98           & 0.98          \\
                                     & \textit{\textbf{Bayesian Network}}     & 84.9           & 0.9          & 0.97           & 0.85          \\
                                     & \textit{\textbf{Naive Bayes}}          & 76.58          & 0.85         & 0.97           & 0.77          \\ \cmidrule(r){1-1}
\multirow{7}{*}{\textbf{3}}          & \textit{\textbf{DNN (GPT-2 TL 10000)}} & \textbf{99.47} & \textbf{0.99}   & \textbf{0.99}     & \textbf{0.99}    \\
                                     & \textit{\textbf{DNN (LSTM TL 2500)}}   & 99.2           & 0.99         & 0.99           & 0.99          \\
                                     & \textit{\textbf{SMO}}                  & 99.15          & 0.99         & 0.99           & 0.98          \\
                                     & \textit{\textbf{Logistic Regression}}  & 98.85          & 0.99         & 0.99           & 0.98          \\
                                     & \textit{\textbf{Random Forest}}        & 98.79          & 0.99         & 0.99           & 0.98          \\
                                     & \textit{\textbf{Bayesian Network}}     & 91.49          & 0.94         & 0.98           & 0.92          \\
                                     & \textit{\textbf{Naive Bayes}}          & 74.37          & 0.83         & 0.96           & 0.74          \\ \cmidrule(r){1-1}
\multirow{7}{*}{\textbf{4}}          & \textit{\textbf{DNN (GPT-2 TL 10000)}} & \textbf{99.56} & \textbf{$\sim$1}   & \textbf{$\sim$1}     & \textbf{$\sim$1}    \\
                                     & \textit{\textbf{DNN (LSTM TL 5000)}}   & 99.22          & 0.99         & 0.99           & 0.99          \\
                                     & \textit{\textbf{Logistic Regression}}  & 98.66          & 0.99         & 0.98           & 0.98          \\
                                     & \textit{\textbf{SMO}}                  & 98.66          & 0.99         & 0.98           & 0.98          \\
                                     & \textit{\textbf{Random Forest}}        & 98             & 0.98         & 0.98           & 0.98          \\
                                     & \textit{\textbf{Bayesian Network}}     & 95.53          & 0.96         & 0.98           & 0.96          \\
                                     & \textit{\textbf{Naive Bayes}}          & 88.74          & 0.92         & 0.97           & 0.89          \\ \cmidrule(r){1-1}
\multirow{7}{*}{\textbf{5}}          & \textit{\textbf{DNN (GPT-2 TL 10000)}} & \textbf{99.85} & \textbf{$\sim$1}   & \textbf{$\sim$1}     & \textbf{$\sim$1}    \\
                                     & \textit{\textbf{DNN (LSTM TL 10000)}}  & 99.67          & 1            & 1              & 1             \\
                                     & \textit{\textbf{Logistic Regression}}  & 98.86          & 0.99         & 0.99           & 0.99          \\
                                     & \textit{\textbf{Random Forest}}        & 98.7           & 0.99         & 0.99           & 0.99          \\
                                     & \textit{\textbf{SMO}}                  & 98.6           & 0.99         & 0.99           & 0.99          \\
                                     & \textit{\textbf{Naive Bayes}}          & 90.55          & 0.94         & 0.98           & 0.9           \\
                                     & \textit{\textbf{Bayesian Network}}     & 88.95          & 0.93         & 0.98           & 0.89          \\ \cmidrule(r){1-1}
\multirow{7}{*}{\textbf{6}}          & \textit{\textbf{DNN (GPT-2 TL 7500)}}  & \textbf{99.82} & \textbf{$\sim$1}   & \textbf{$\sim$1}     & \textbf{$\sim$1}    \\
                                     & \textit{\textbf{DNN (LSTM TL 2500)}}   & 99.8           & $\sim$1            & $\sim$1              & $\sim$1             \\
                                     & \textit{\textbf{Logistic Regression}}  & 99.1           & 0.99         & 0.99           & 0.99          \\
                                     & \textit{\textbf{Random Forest}}        & 98.9           & 0.99         & 0.99           & 0.99          \\
                                     & \textit{\textbf{SMO}}                  & 98.86          & 0.99         & 0.99           & 0.99          \\
                                     & \textit{\textbf{Naive Bayes}}          & 90.52          & 0.94         & 0.98           & 0.9           \\
                                     & \textit{\textbf{Bayesian Network}}     & 89.27          & 0.93         & 0.98           & 0.89          \\ \cmidrule(r){1-1}
\multirow{7}{*}{\textbf{7}}          & \textit{\textbf{DNN (LSTM TL 10000)}}  & \textbf{99.91} & \textbf{$\sim$1}   & \textbf{$\sim$1}     & \textbf{$\sim$1}    \\
                                     & \textit{\textbf{DNN (GPT-2 TL 5000)}}  & 99.86          & $\sim$1            & $\sim$1              & $\sim$1             \\
                                     & \textit{\textbf{SMO}}                  & 99.4           & 0.99         & 0.99           & 0.99          \\
                                     & \textit{\textbf{Logistic Regression}}  & 99.13          & 0.99         & 0.99           & 0.99          \\
                                     & \textit{\textbf{Random Forest}}        & 99             & 0.99         & 0.99           & 0.99          \\
                                     & \textit{\textbf{Bayesian Network}}     & 88.67          & 0.93         & 0.98           & 0.89          \\
                                     & \textit{\textbf{Naive Bayes}}          & 86.9           & 0.91         & 0.98           & 0.87          \\ \bottomrule
\end{tabular}
\end{table}

\begin{table}[]
\caption{Average performance of the chosen models for each of the 7 subjects.}
\label{tab:average-comparison}
\centering
\begin{tabular}{@{}lllll@{}}
\toprule
\textbf{Model}                        & \textbf{Avg acc} & \textbf{F-1} & \textbf{Prec.} & \textbf{Rec.} \\ \midrule
\textit{\textbf{DNN (LSTM TL)}}       & 99.57            & $\sim$1      & $\sim$1        & $\sim$1       \\
\textit{\textbf{DNN (GPT-2 TL)}}      & 99.43            & $\sim$1      & $\sim$1        & $\sim$1       \\
\textit{\textbf{SMO}}                 & 98.76            & 0.99         & 0.98           & 0.98          \\
\textit{\textbf{Logistic Regression}} & 98.63            & 0.99         & 0.98           & 0.98          \\
\textit{\textbf{Random Forest}}       & 98.45            & 0.98         & 0.98           & 0.98          \\
\textit{\textbf{Bayesian Network}}    & 88.73            & 0.92         & 0.98           & 0.89          \\
\textit{\textbf{Naive Bayes}}         & 83.80            & 0.89         & 0.97           & 0.83          \\ \bottomrule
\end{tabular}
\end{table}

Table \ref{tab:bigcomparison} shows a comparison of the models proposed in this paper to other state-of-the-art methods of speaker recognition. Namely, they are Sequential Minimal Optimisation (SMO), Logistic Regression, Bayesian Networks, and Naive Bayes. It can be observed that, although in some cases close, the DNN fine tuned from synthetic data generated by both the LSTM and GPT-2 achieve higher scores than other methods. Finally, Table \ref{tab:average-comparison} shows the average scores for the chosen models for each of the seven subjects.

\section{Conclusion and Future Work}
\label{sec4:futureworkconclusion}
To finally conclude, this work found strong success for all 7 subjects when improving the classification problem of speaker recognition by generating augmented data by both LSTM and OpenAI GPT-2 models. Future work aims to solidify this hypothesis by running the experiments for a large range of subjects and comparing the patterns that emerge from the results. 

The experiments in this work have provided a strong argument for the usage of deep neural network transfer learning from MFCCs synthesised by both LSTM and GPT-2 models for the problem of speaker recognition. One of the limitations of this study was hardware availability since it was focused on those available to consumers today. The Flickr8k dataset was thus limited to 8,000 data objects and new datasets created, which prevents a direct comparison to other speaker recognition works which often operate on larger data and with hardware beyond consumer availability. It is worth noting that the complex nature of training LSTM and GPT-2 models to generate MFCCs is beyond that of the task of speaker recognition itself, and as such, devices with access to TPU or CUDA-based hardware must perform the task in the background over time. The tasks in question took several minutes with the two GPUs used in this work for both LSTM and GPT-2 and as such are not instantaneous. As previously mentioned, although it was observed that overfitting did not occur too strongly, it would be useful in future to perform similar experiments with either K-fold or leave-one-out Cross Validation in order to achieve even more accurate representations of the classification metrics. In terms of future application, the then-optimised model would be the implemented within real robots and smarthome assistants through compatible software.

In this work, seven subjects were benchmarked with both a tuned LSTM and OpenAI's GPT-2 model. In future, as was seen with related works, a GAN could also be implemented in order to provide a third possible solution to the problem of data scarcity in speaker recognition as well as other related speech recognition classification problems - bias is an open issue that has been noted for data augmentation with GANs~\cite{hu2019exploring}, and as such, this issue must be studied if a GAN is implemented for problems of this nature. This work further argued for the hypothesis presented, that is, data augmentation can aid in improving speaker recognition for scarce datasets. Following the 14 successful runs including LSTMs and GPT-2s, the overall process followed by this experiment could be scripted and thus completely automated, allowing for the benchmarking of many more subjects to give a much more generalised set of results, of which will be more representative of the general population. Additionally, samples spoken from many languages should also be considered in order to provide language generalisation, rather than just the English language spoken by multiple international dialects in this study. Should generalisation be possible, future models may require only a small amount of fine-tuning to produce synthetic speech for a given person rather than training from scratch as was performed in this work. In more general lines of thought, literature review shows that much of the prominent work is recent, leaving many fields of machine learning that have not yet been attempted to be improved via the methods described in this work. 

\bibliographystyle{ieeetr} 
\bibliography{bibliography} 

\begin{thebibliography}{10}

\bibitem{logsdon2018alexa}
A.~Logsdon~Smith, ``Alexa, who owns my pillow talk? contracting, collaterizing,
  and monetizing consumer privacy through voice-captured personal data,'' {\em
  Catholic University Journal of Law and Technology}, vol.~27, no.~1,
  pp.~187--226, 2018.

\bibitem{dunin2020alexa}
A.~Dunin-Underwood, ``Alexa, can you keep a secret? applicability of the
  third-party doctrine to information collected in the home by virtual
  assistants,'' {\em Information \& Communications Technology Law}, pp.~1--19,
  2020.

\bibitem{babbar2019data}
R.~Babbar and B.~Sch{\"o}lkopf, ``Data scarcity, robustness and extreme
  multi-label classification,'' {\em Machine Learning}, vol.~108, no.~8-9,
  pp.~1329--1351, 2019.

\bibitem{wang2017effectiveness}
J.~Wang and L.~Perez, ``The effectiveness of data augmentation in image
  classification using deep learning,'' {\em Convolutional Neural Networks Vis.
  Recognit}, p.~11, 2017.

\bibitem{zhu2018emotion}
X.~Zhu, Y.~Liu, J.~Li, T.~Wan, and Z.~Qin, ``Emotion classification with data
  augmentation using generative adversarial networks,'' in {\em Pacific-Asia
  Conference on Knowledge Discovery and Data Mining}, pp.~349--360, Springer,
  2018.

\bibitem{frid2018synthetic}
M.~Frid-Adar, E.~Klang, M.~Amitai, J.~Goldberger, and H.~Greenspan, ``Synthetic
  data augmentation using gan for improved liver lesion classification,'' in
  {\em 2018 IEEE 15th international symposium on biomedical imaging (ISBI
  2018)}, pp.~289--293, IEEE, 2018.

\bibitem{yang2018se}
J.~H. Yang, N.~K. Kim, and H.~K. Kim, ``Se-resnet with gan-based data
  augmentation applied to acoustic scene classification,'' in {\em DCASE 2018
  workshop}, 2018.

\bibitem{madhu2019data}
A.~Madhu and S.~Kumaraswamy, ``Data augmentation using generative adversarial
  network for environmental sound classification,'' in {\em 2019 27th European
  Signal Processing Conference (EUSIPCO)}, pp.~1--5, IEEE, 2019.

\bibitem{bird2020overcoming}
J.~J. Bird, D.~R. Faria, C.~Premebida, A.~Ek{\'a}rt, and P.~P. Ayrosa,
  ``Overcoming data scarcity in speaker identification: Dataset augmentation
  with synthetic mfccs via character-level rnn,'' in {\em 2020 IEEE
  International Conference on Autonomous Robot Systems and Competitions
  (ICARSC)}, pp.~146--151, IEEE, 2020.

\bibitem{poddar2017speaker}
A.~Poddar, M.~Sahidullah, and G.~Saha, ``Speaker verification with short
  utterances: a review of challenges, trends and opportunities,'' {\em IET
  Biometrics}, vol.~7, no.~2, pp.~91--101, 2017.

\bibitem{mumolo2003distant}
E.~Mumolo and M.~Nolich, ``Distant talker identification by nonlinear
  programming and beamforming in service robotics,'' in {\em IEEE-EURASIP
  Workshop on Nonlinear Signal and Image Processing}, pp.~8--11, 2003.

\bibitem{ratha2001automated}
N.~K. Ratha, A.~Senior, and R.~M. Bolle, ``Automated biometrics,'' in {\em
  International Conference on Advances in Pattern Recognition}, pp.~447--455,
  Springer, 2001.

\bibitem{rose2002forensic}
P.~Rose, {\em Forensic speaker identification}.
\newblock cRc Press, 2002.

\bibitem{hasan2004speaker}
M.~R. Hasan, M.~Jamil, M.~Rahman, {\em et~al.}, ``Speaker identification using
  mel frequency cepstral coefficients,'' {\em variations}, vol.~1, no.~4, 2004.

\bibitem{nagrani2017voxceleb}
A.~Nagrani, J.~S. Chung, and A.~Zisserman, ``Voxceleb: a large-scale speaker
  identification dataset,'' {\em arXiv preprint arXiv:1706.08612}, 2017.

\bibitem{yadav2018learning}
S.~Yadav and A.~Rai, ``Learning discriminative features for speaker
  identification and verification.,'' in {\em Interspeech}, pp.~2237--2241,
  2018.

\bibitem{zeinali2019but}
H.~Zeinali, S.~Wang, A.~Silnova, P.~Mat{\v{e}}jka, and O.~Plchot, ``But system
  description to voxceleb speaker recognition challenge 2019,'' {\em arXiv
  preprint arXiv:1910.12592}, 2019.

\bibitem{graves2013hybrid}
A.~Graves, N.~Jaitly, and A.-r. Mohamed, ``Hybrid speech recognition with deep
  bidirectional lstm,'' in {\em Automatic Speech Recognition and Understanding
  (ASRU), 2013 IEEE Workshop on}, pp.~273--278, IEEE, 2013.

\bibitem{belin2000voice}
P.~Belin, R.~J. Zatorre, P.~Lafaille, P.~Ahad, and B.~Pike, ``Voice-selective
  areas in human auditory cortex,'' {\em Nature}, vol.~403, no.~6767,
  pp.~309--312, 2000.

\bibitem{radford2018improving}
A.~Radford, K.~Narasimhan, T.~Salimans, and I.~Sutskever, ``Improving language
  understanding by generative pre-training,'' {\em URL https://s3-us-west-2.
  amazonaws. com/openai-assets/researchcovers/languageunsupervised/language
  understanding paper. pdf}, 2018.

\bibitem{radford2019language}
A.~Radford, J.~Wu, R.~Child, D.~Luan, D.~Amodei, and I.~Sutskever, ``Language
  models are unsupervised multitask learners,'' {\em OpenAI Blog}, vol.~1,
  no.~8, p.~9, 2019.

\bibitem{jelinek1980interpolated}
F.~Jelinek, ``Interpolated estimation of markov source parameters from sparse
  data,'' in {\em Proc. Workshop on Pattern Recognition in Practice, 1980},
  1980.

\bibitem{bengio2003neural}
Y.~Bengio, R.~Ducharme, P.~Vincent, and C.~Jauvin, ``A neural probabilistic
  language model,'' {\em Journal of machine learning research}, vol.~3,
  no.~Feb, pp.~1137--1155, 2003.

\bibitem{vaswani2017attention}
A.~Vaswani, N.~Shazeer, N.~Parmar, J.~Uszkoreit, L.~Jones, A.~N. Gomez,
  {\L}.~Kaiser, and I.~Polosukhin, ``Attention is all you need,'' in {\em
  Advances in neural information processing systems}, pp.~5998--6008, 2017.

\bibitem{solaiman2019release}
I.~Solaiman, M.~Brundage, J.~Clark, A.~Askell, A.~Herbert-Voss, J.~Wu,
  A.~Radford, and J.~Wang, ``Release strategies and the social impacts of
  language models,'' {\em arXiv preprint arXiv:1908.09203}, 2019.

\bibitem{gehrmann2019gltr}
S.~Gehrmann, H.~Strobelt, and A.~M. Rush, ``Gltr: Statistical detection and
  visualization of generated text,'' {\em arXiv preprint arXiv:1906.04043},
  2019.

\bibitem{wolff2020attacking}
M.~Wolff, ``Attacking neural text detectors,'' {\em arXiv preprint
  arXiv:2002.11768}, 2020.

\bibitem{adelani2020generating}
D.~I. Adelani, H.~Mai, F.~Fang, H.~H. Nguyen, J.~Yamagishi, and I.~Echizen,
  ``Generating sentiment-preserving fake online reviews using neural language
  models and their human-and machine-based detection,'' in {\em International
  Conference on Advanced Information Networking and Applications},
  pp.~1341--1354, Springer, 2020.

\bibitem{beres1960perception}
D.~Beres, ``Perception, imagination, and reality,'' {\em International Journal
  of Psycho-Analysis}, vol.~41, pp.~327--334, 1960.

\bibitem{egan1989memory}
K.~Egan, ``Memory, imagination, and learning: Connected by the story,'' {\em
  Phi Delta Kappan}, vol.~70, no.~6, pp.~455--459, 1989.

\bibitem{heath2008exploring}
G.~Heath, ``Exploring the imagination to establish frameworks for learning,''
  {\em Studies in Philosophy and Education}, vol.~27, no.~2-3, pp.~115--123,
  2008.

\bibitem{macintyre2012emotions}
P.~MacIntyre and T.~Gregersen, ``Emotions that facilitate language learning:
  The positive-broadening power of the imagination,'' 2012.

\bibitem{egan2014imagination}
K.~Egan, {\em Imagination in teaching and learning: The middle school years}.
\newblock University of Chicago Press, 2014.

\bibitem{xu2016improved}
Y.~Xu, R.~Jia, L.~Mou, G.~Li, Y.~Chen, Y.~Lu, and Z.~Jin, ``Improved relation
  classification by deep recurrent neural networks with data augmentation,''
  {\em arXiv preprint arXiv:1601.03651}, 2016.

\bibitem{shin2018medical}
H.-C. Shin, N.~A. Tenenholtz, J.~K. Rogers, C.~G. Schwarz, M.~L. Senjem, J.~L.
  Gunter, K.~P. Andriole, and M.~Michalski, ``Medical image synthesis for data
  augmentation and anonymization using generative adversarial networks,'' in
  {\em International workshop on simulation and synthesis in medical imaging},
  pp.~1--11, Springer, 2018.

\bibitem{kobayashi2018contextual}
S.~Kobayashi, ``Contextual augmentation: Data augmentation by words with
  paradigmatic relations,'' {\em arXiv preprint arXiv:1805.06201}, 2018.

\bibitem{zhang2019dada}
X.~Zhang, Z.~Wang, D.~Liu, and Q.~Ling, ``Dada: Deep adversarial data
  augmentation for extremely low data regime classification,'' in {\em ICASSP
  2019-2019 IEEE International Conference on Acoustics, Speech and Signal
  Processing (ICASSP)}, pp.~2807--2811, IEEE, 2019.

\bibitem{barz2020deep}
B.~Barz and J.~Denzler, ``Deep learning on small datasets without pre-training
  using cosine loss,'' in {\em The IEEE Winter Conference on Applications of
  Computer Vision}, pp.~1371--1380, 2020.

\bibitem{huang2018multimodal}
J.~Huang, Y.~Li, J.~Tao, Z.~Lian, M.~Niu, and M.~Yang, ``Multimodal continuous
  emotion recognition with data augmentation using recurrent neural networks,''
  in {\em Proceedings of the 2018 on Audio/Visual Emotion Challenge and
  Workshop}, pp.~57--64, 2018.

\bibitem{tang2018digital}
B.~Tang, Y.~Tu, Z.~Zhang, and Y.~Lin, ``Digital signal modulation
  classification with data augmentation using generative adversarial nets in
  cognitive radio networks,'' {\em IEEE Access}, vol.~6, pp.~15713--15722,
  2018.

\bibitem{park2019specaugment}
D.~S. Park, W.~Chan, Y.~Zhang, C.-C. Chiu, B.~Zoph, E.~D. Cubuk, and Q.~V. Le,
  ``Specaugment: A simple data augmentation method for automatic speech
  recognition,'' {\em arXiv preprint arXiv:1904.08779}, 2019.

\bibitem{wang2017tacotron}
Y.~Wang, R.~Skerry-Ryan, D.~Stanton, Y.~Wu, R.~J. Weiss, N.~Jaitly, Z.~Yang,
  Y.~Xiao, Z.~Chen, S.~Bengio, {\em et~al.}, ``Tacotron: Towards end-to-end
  speech synthesis,'' {\em arXiv preprint arXiv:1703.10135}, 2017.

\bibitem{chien2018adversarial}
J.-T. Chien and K.-T. Peng, ``Adversarial learning and augmentation for speaker
  recognition.,'' in {\em Odyssey}, pp.~342--348, 2018.

\bibitem{eck2002finding}
D.~Eck and J.~Schmidhuber, ``Finding temporal structure in music: Blues
  improvisation with lstm recurrent networks,'' in {\em Proceedings of the 12th
  IEEE workshop on neural networks for signal processing}, pp.~747--756, IEEE,
  2002.

\bibitem{senjyu2006application}
T.~Senjyu, A.~Yona, N.~Urasaki, and T.~Funabashi, ``Application of recurrent
  neural network to long-term-ahead generating power forecasting for wind power
  generator,'' in {\em 2006 IEEE PES Power Systems Conference and Exposition},
  pp.~1260--1265, IEEE, 2006.

\bibitem{pawade2018story}
D.~Pawade, A.~Sakhapara, M.~Jain, N.~Jain, and K.~Gada, ``Story
  scrambler--automatic text generation using word level rnn-lstm,'' {\em
  International Journal of Information Technology and Computer Science
  (IJITCS)}, vol.~10, no.~6, pp.~44--53, 2018.

\bibitem{sha2018order}
L.~Sha, L.~Mou, T.~Liu, P.~Poupart, S.~Li, B.~Chang, and Z.~Sui,
  ``Order-planning neural text generation from structured data,'' in {\em
  Thirty-Second AAAI Conference on Artificial Intelligence}, 2018.

\bibitem{gregor2015draw}
K.~Gregor, I.~Danihelka, A.~Graves, D.~J. Rezende, and D.~Wierstra, ``Draw: A
  recurrent neural network for image generation,'' {\em arXiv preprint
  arXiv:1502.04623}, 2015.

\bibitem{wang2017rnn}
X.~Wang, S.~Takaki, and J.~Yamagishi, ``An rnn-based quantized f0 model with
  multi-tier feedback links for text-to-speech synthesis.,'' in {\em
  INTERSPEECH}, pp.~1059--1063, 2017.

\bibitem{fernandez2007application}
S.~Fern{\'a}ndez, A.~Graves, and J.~Schmidhuber, ``An application of recurrent
  neural networks to discriminative keyword spotting,'' in {\em International
  Conference on Artificial Neural Networks}, pp.~220--229, Springer, 2007.

\bibitem{he2019streaming}
Y.~He, T.~N. Sainath, R.~Prabhavalkar, I.~McGraw, R.~Alvarez, D.~Zhao,
  D.~Rybach, A.~Kannan, Y.~Wu, R.~Pang, {\em et~al.}, ``Streaming end-to-end
  speech recognition for mobile devices,'' in {\em ICASSP 2019-2019 IEEE
  International Conference on Acoustics, Speech and Signal Processing
  (ICASSP)}, pp.~6381--6385, IEEE, 2019.

\bibitem{sak2014long}
H.~Sak, A.~W. Senior, and F.~Beaufays, ``Long short-term memory recurrent
  neural network architectures for large scale acoustic modeling,'' 2014.

\bibitem{valentini2016investigating}
C.~Valentini-Botinhao, X.~Wang, S.~Takaki, and J.~Yamagishi, ``Investigating
  rnn-based speech enhancement methods for noise-robust text-to-speech.,'' in
  {\em SSW}, pp.~146--152, 2016.

\bibitem{tachibana2018efficiently}
H.~Tachibana, K.~Uenoyama, and S.~Aihara, ``Efficiently trainable
  text-to-speech system based on deep convolutional networks with guided
  attention,'' in {\em 2018 IEEE International Conference on Acoustics, Speech
  and Signal Processing (ICASSP)}, pp.~4784--4788, IEEE, 2018.

\bibitem{harwath2015deep}
D.~Harwath and J.~Glass, ``Deep multimodal semantic embeddings for speech and
  images,'' in {\em 2015 IEEE Workshop on Automatic Speech Recognition and
  Understanding (ASRU)}, pp.~237--244, IEEE, 2015.

\bibitem{rothauser1969ieee}
E.~Rothauser, ``Ieee recommended practice for speech quality measurements,''
  {\em IEEE Trans. on Audio and Electroacoustics}, vol.~17, pp.~225--246, 1969.

\bibitem{bird2019phoneme2}
J.~J. Bird, A.~Ek{\'a}rt, and D.~R. Faria, ``Phoneme aware speech synthesis via
  fine tune transfer learning with a tacotron spectrogram prediction network,''
  in {\em UK Workshop on Computational Intelligence}, pp.~271--282, Springer,
  2019.

\bibitem{xiong2003comparing}
Z.~Xiong, R.~Radhakrishnan, A.~Divakaran, and T.~S. Huang, ``Comparing {MFCC}
  and mpeg-7 audio features for feature extraction, maximum likelihood {HMM}
  and entropic prior {HMM} for sports audio classification,'' in {\em IEEE Int.
  Conference on Acoustics, Speech, and Signal Processing (ICASSP).}, vol.~5,
  2003.

\bibitem{bird2019phoneme}
J.~J. Bird, E.~Wanner, A.~Ek{\'a}rt, and D.~R. Faria, ``Phoneme aware speech
  recognition through evolutionary optimisation,'' in {\em Proceedings of the
  Genetic and Evolutionary Computation Conference Companion}, pp.~362--363,
  2019.

\bibitem{muda2010voice}
L.~Muda, M.~Begam, and I.~Elamvazuthi, ``Voice recognition algorithms using mel
  frequency cepstral coefficient ({MFCC}) and dynamic time warping ({DTW})
  techniques,'' {\em arXiv preprint arXiv:1003.4083}, 2010.

\bibitem{sahidullah2012design}
M.~Sahidullah and G.~Saha, ``Design, analysis and experimental evaluation of
  block based transformation in mfcc computation for speaker recognition,''
  {\em Speech Communication}, vol.~54, no.~4, pp.~543--565, 2012.

\bibitem{stevens1937scale}
S.~S. Stevens, J.~Volkmann, and E.~B. Newman, ``A scale for the measurement of
  the psychological magnitude pitch,'' {\em The Journal of the Acoustical
  Society of America}, vol.~8, no.~3, pp.~185--190, 1937.

\bibitem{kingma2014adam}
D.~P. Kingma and J.~Ba, ``Adam: A method for stochastic optimization,'' {\em
  arXiv preprint arXiv:1412.6980}, 2014.

\bibitem{bird2020optimisation}
J.~J. Bird, E.~Wanner, A.~Ek{\'a}rt, and D.~R. Faria, ``Optimisation of
  phonetic aware speech recognition through multi-objective evolutionary
  algorithms,'' {\em Expert Systems with Applications}, p.~113402, 2020.

\bibitem{hu2019exploring}
M.~Hu and J.~Li, ``Exploring bias in gan-based data augmentation for small
  samples,'' {\em arXiv preprint arXiv:1905.08495}, 2019.

\end{thebibliography}

%


\begin{IEEEbiography}[{\includegraphics[width=1in,height=1.25in,clip,keepaspectratio]{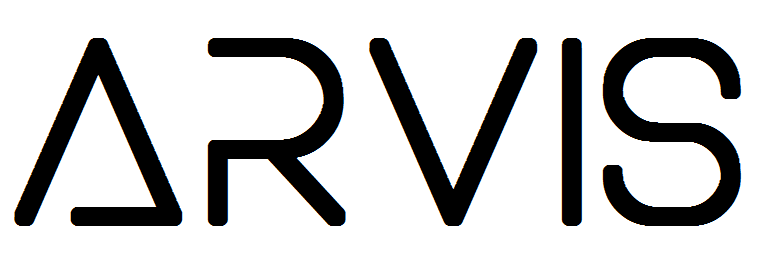}}]{Aston Robotics, Vision and Intelligent Systems (ARVIS) Lab}
aims to improve mankind’s quality of life by enabling intelligent robots, virtual agents and autonomous systems with the perceptual and cognitive capabilities of the future. The team is made up of researchers from the Computer Science Department and is within ASTUTE (Aston Institute for Urban Technology and the Environment), at the School of Engineering and Applied Science of Aston University, Birmingham, UK.\\ \textbf{https://arvis-lab.io/}
\end{IEEEbiography}

\end{document}